\documentclass[a4paper,USenglish,cleveref,autoref,thm-restate]{lipics-v2021}

\usepackage{microtype}   
\usepackage{hyperref}
\usepackage[numbers]{natbib}
\usepackage[inline]{enumitem}
\usepackage{spotltl}
\usepackage{listings}

\lstdefinestyle{ShStyle}{
  language=bash,
  basicstyle=\small\ttfamily,
  frame=l,
  columns=fullflexible,
  backgroundcolor=\color{yellow!20},
  aboveskip=2pt,
  belowskip=2pt,
}
\lstdefinestyle{ShStyleOut}{
  language=bash,
  basicstyle=\small\ttfamily,
  frame=l,
  columns=fullflexible,
  backgroundcolor=\color{yellow!10},
  aboveskip=2pt,
  belowskip=2pt,
}

\title{Teaching LTL and \texorpdfstring{$\omega$}{omega}-Automata with Spot}

\author{Alexandre Duret-Lutz}
       {EPITA Research Laboratory (LRE), Paris, France}
       {adl@lre.epita.fr}
       {https://orcid.org/0000-0002-6623-2512}
       {}

\authorrunning{A. Duret-Lutz} 

\Copyright{Alexandre Duret-Lutz}



\category{Demo} 

\relatedversion{}   

\supplement{
  Online tool: \url{https://spot.lre.epita.fr/app/}\\
  Interactive notebooks: \url{https://spot.lre.epita.fr/tut.html}\\
  Source code: \url{https://gitlab.lre.epita.fr/spot/spot}
}

\nolinenumbers  

\begin{document}

\maketitle

\begin{abstract}
  Spot is a mature, open-source C++/Python library and toolset for
  Linear Temporal Logic (LTL) and $\omega$-automata manipulation.
  While Spot is routinely used as a research and verification
  back-end, its rich visualization capabilities and Python interface
  also make it an attractive platform for \emph{teaching} the
  connections between temporal logic formulas and the
  $\omega$-automata that give them their semantics.

This demonstration showcases two complementary, zero-install entry points
into Spot that are suitable for educational settings:
\begin{enumerate*}
\item a web application that lets students type LTL or PSL formulas
  and immediately see the resulting automaton, explore formula
  simplifications, compare formula equivalence/implication, and
  navigate Manna \& Pnueli's temporal hierarchy.
\item Jupyter notebooks that combine narrative explanations, live
  Python code, and inline automaton drawings, enabling students to
  experiment interactively and instructors to build assignments around
  concrete, executable examples.
\end{enumerate*}
Additionally, we can discuss the use of the command-line tools of Spot
to generate random examples suitable for preparing a series of exercises.
\end{abstract}

\section{LTL and Spot}
\label{sec:intro}

Linear Temporal Logic (LTL)~\cite{pnueli.77.focs} and
$\omega$-automata are cornerstones of formal methods education.  LTL
appears in virtually every graduate course on program verification,
and the translation of an LTL formula into a Büchi automaton is an
important step to better understand the logic and its many
applications, like model checking~\cite{vardi.07.vmcai} or reactive
synthesis~\cite{piterman.06.vmcai}.

Several textbooks present the theoretical foundations~\cite[e.g.][]{baier.08.book,clarke.99.book}, but understanding deepens
considerably when students can \emph{experiment}: vary a formula, see
the resulting automaton change, and build intuition about which formulas
yield small deterministic automata and which do not.

\textbf{Spot in a nutshell.}
Spot~\cite{duret.16.atva2,duret.22.cav} is an open-source library (GPL v3) for the manipulation
of LTL formulas and $\omega$-automata.     It supports:
\begin{itemize}
  \item LTL and a subset of PSL~\cite{psl.04.lrm}, with multiple input syntaxes;
  \item $\omega$-automata with arbitrary (Emerson--Lei) acceptance
        conditions, in five standard formats (HOA~\cite{babiak.15.cav},
        never claims, LBTT, DSTAR, PGsolver);
      \item a comprehensive set of formula and automaton algorithms:
        simplification, equivalence and implication testing, checking
        stutter-invariance~\cite{michaud.15.spin}, translation to
        automata (generalized Büchi, Rabin, Streett, parity, \ldots),
        simulation-based reduction~\cite{babiak.13.spin} and SAT-based
        minimization~\cite{baarir.15.lpar}, and more.
  \item recent support for LTLf (LTL over finite traces) and DFA represented
    using multi-terminal binary decision diagrams~\cite{duret.25.ciaa}.
\end{itemize}
Beyond the C++ library, Spot exposes all algorithms through
command-line tools and Python bindings, and generates automaton
pictures rendered via Graphviz.

\section{The Online LTL Toolset}
\label{sec:webapp}

\begin{figure}[t]
  \centering
  \begin{minipage}[b]{.495\textwidth}
    \includegraphics[width=\textwidth]{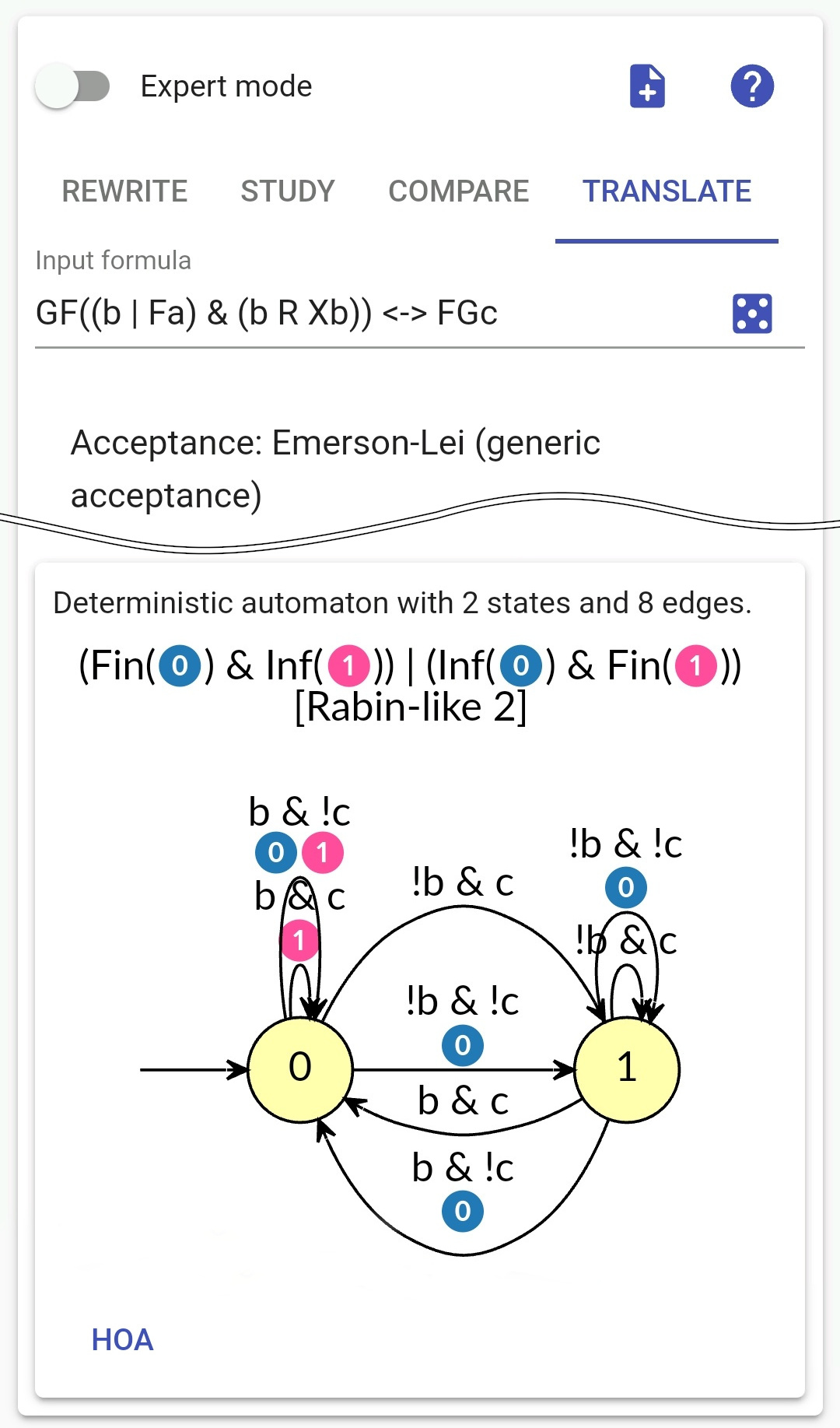}
  \end{minipage}
  \begin{minipage}[b]{.495\textwidth}
    \includegraphics[width=\textwidth]{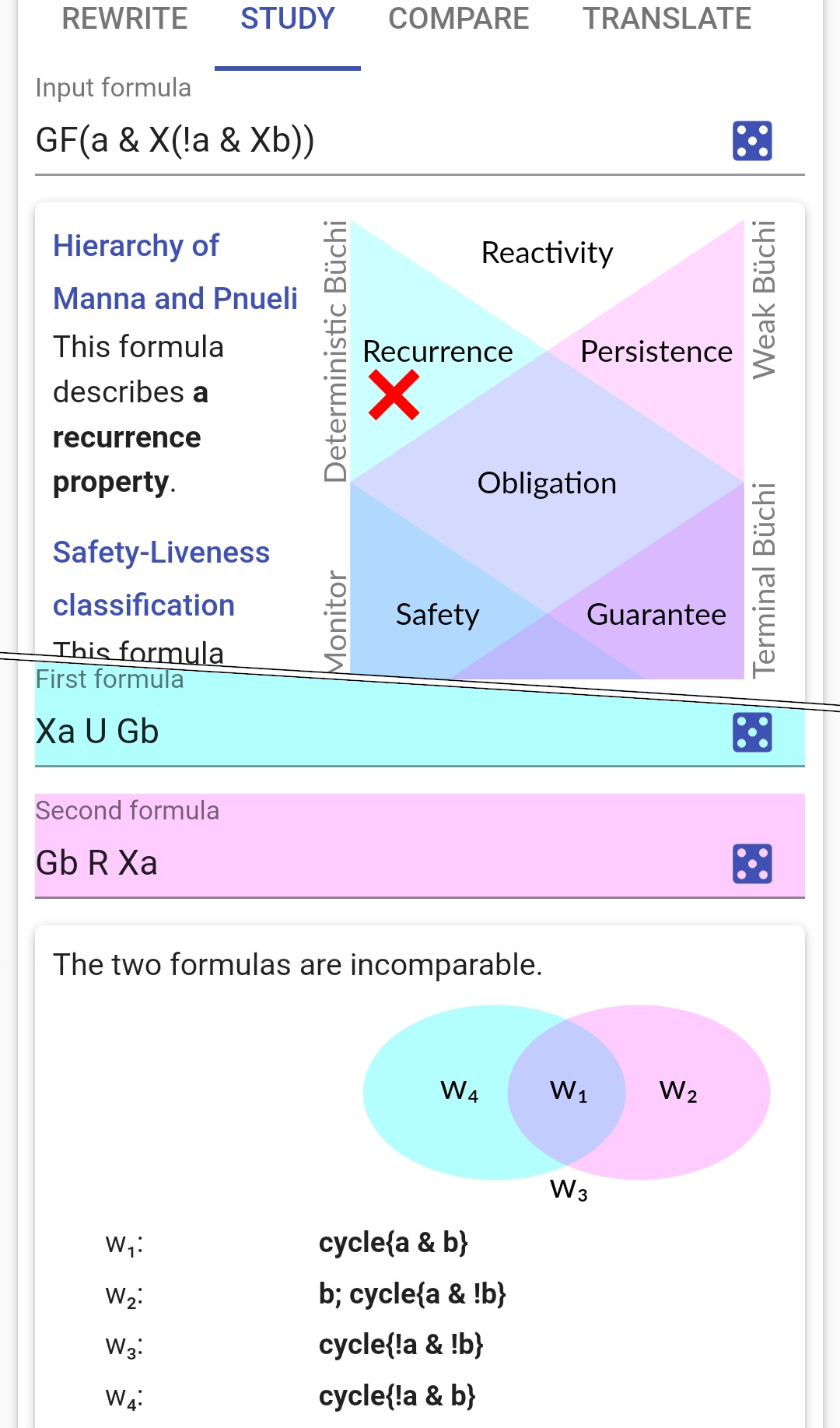}
  \end{minipage}
  \caption[Spot's online LTL toolset]{Spot's online LTL toolset~\cite{duret.22.cav}.}
  \label{fig:webapp}
\end{figure}

Spot's online LTL toolset, available at
\url{https://spot.lre.epita.fr/app/}, is a single-page web application
that requires no installation.  Figure~\ref{fig:webapp} shows a collage
of three different interfaces:
\begin{itemize}
\item The left is an example of the \textsc{translate} tab, where an LTL
  formula can be converted to $\omega$-automata with various
  acceptance conditions.
\item The \textsc{study} tab in the upper right computes various
  characteristics of LTL formulas.
\item The bottom right example shows the \textsc{compare} tab, where
  two formulas can be compared, returning distinguishing
  $\omega$-words when they exist.
\end{itemize}
A fourth tab, not illustrated here, is dedicated to the simplification
of formulas.

This tool can be used by students (but also researchers) to improve
their understanding of LTL, PSL, and $\omega$-automata.

Because the tool runs in a standard web browser, it is suitable for
classroom use on any device, including tablets.  Internet access is
required to query the server, where Spot is installed.

The source code for the client and the server of this web application
can be found at \url{https://gitlab.lre.epita.fr/spot/spot-web-app/}
for anyone willing to run it locally, or extend it.

Here are a few examples of activities that this tool can be used for:
\begin{itemize}
\item Translating a temporal logic formula to an $\omega$-automaton is
  of course a way to understand the semantics of a given formula.
\item Comparing two formulas, and showing examples of words that
  distinguish them, is a great way to work on finding equivalent
  formulas.  For instance, if a student is tasked with finding
  a simpler formula equivalent to $(a\R(a \U b))\W b$, the student
  could translate this formula into an automaton, and recognize
  that this looks like the automaton for $a \U b$ and use the
  \textsc{compare} tab to verify.  Or the student could play with
  the formula in the \textsc{compare} tab, and discover that
  $(a\R(a \U b))\W b \equiv (a \U b)\W b \equiv a \U b$.
\item Being able to choose between multiple acceptance conditions in the
  \textsc{translate} tab can help illustrate the expressiveness and
  succinctness of the different types of $\omega$-automata.
\item These different types of automata can also be related to the
  classical hierarchy of Manna and Pnueli~\cite{manna.87.podc}.  For
  any given formula, the \textsc{study} tab will point to the most precise
  class the formula belongs to, and this can be used to select the most
  appropriate type of $\omega$-automaton to build.
\item The syntactic-future hierarchy is the syntactic counterpart of
  the Manna and Pnueli hierarchy~\cite[Fig.~1]{esparza.24.acm}.  It
  classifies LTL formulas according to their nesting of strong
  operators ($\F$, $\U$, $\M$) and weak operators ($\G$, $\W$, $\R$).
  The two hierarchies are related in the sense that if an LTL property
  belongs to some class of the Manna-Pnueli hierarchy, it can be
  represented by a formula in the corresponding class of the
  syntactic-future hierarchy.  Sometimes, you can consider a formula
  such as $a \W \F(b)$ that is an Obligation property in the
  Manna-Pnueli hierarchy (corresponding to the syntactic class
  $\Delta_1$) but that syntactically belongs to class $\Pi_2$ (which
  would be the Recurrence class of the Manna-Pnueli hierarchy).  The
  tool points out that the discrepancy indicates that there exists an
  equivalent formula in class $\Delta_1$ (where strong and weak
  operators may not be nested).  Playing with the \textsc{compare} tab
  can help figure out that $a \W \F(b) \equiv \G(a) \lor \F(b)$.
\item A property of LTL formulas that is interesting from the point of
  view of model checking is that of ``\emph{stutter
  invariance}''~\cite{etessami.00.ipl} because it allows model checker
  to apply optimizations known as ``\emph{partial order reductions}'' \cite[Chap.~8]{baier.08.book}
  \cite[Chap.~10]{clarke.99.book}.  A formula $\varphi$ is
  stutter-invariant if an $\omega$-word $w$ is accepted by $\varphi$
  if and only if any $\omega$-word $w'$ derived from $w$ by
  duplicating some letters or removing some duplicate letters is also
  accepted by $\varphi$.  It is well known that $\X$-free formulas are
  stutter-invariant, but this is not a necessary condition.  For
  instance the formula
  $\mathsf{F}(a \land \mathsf{X}(\lnot a\land b))$ is
  stutter-invariant.  The \textsc{study} tab of the tool will tell if
  a formula is stutter-invariant, and when that is not the case, it
  will provide some example words $w$ and $w'$ showing why that is not
  the case~\cite{michaud.15.spin}.
\end{itemize}

\section{Jupyter Notebooks}
\label{sec:notebooks}

For courses where students write code (or for researchers exploring
Spot programmatically), Spot provides a rich Python API together with
a gallery of Jupyter~\cite{jupyter.16.elpub} notebooks at
\url{https://spot.lre.epita.fr/tut.html}.

\begin{figure}[t]
  \centering
  \includegraphics[width=\textwidth]{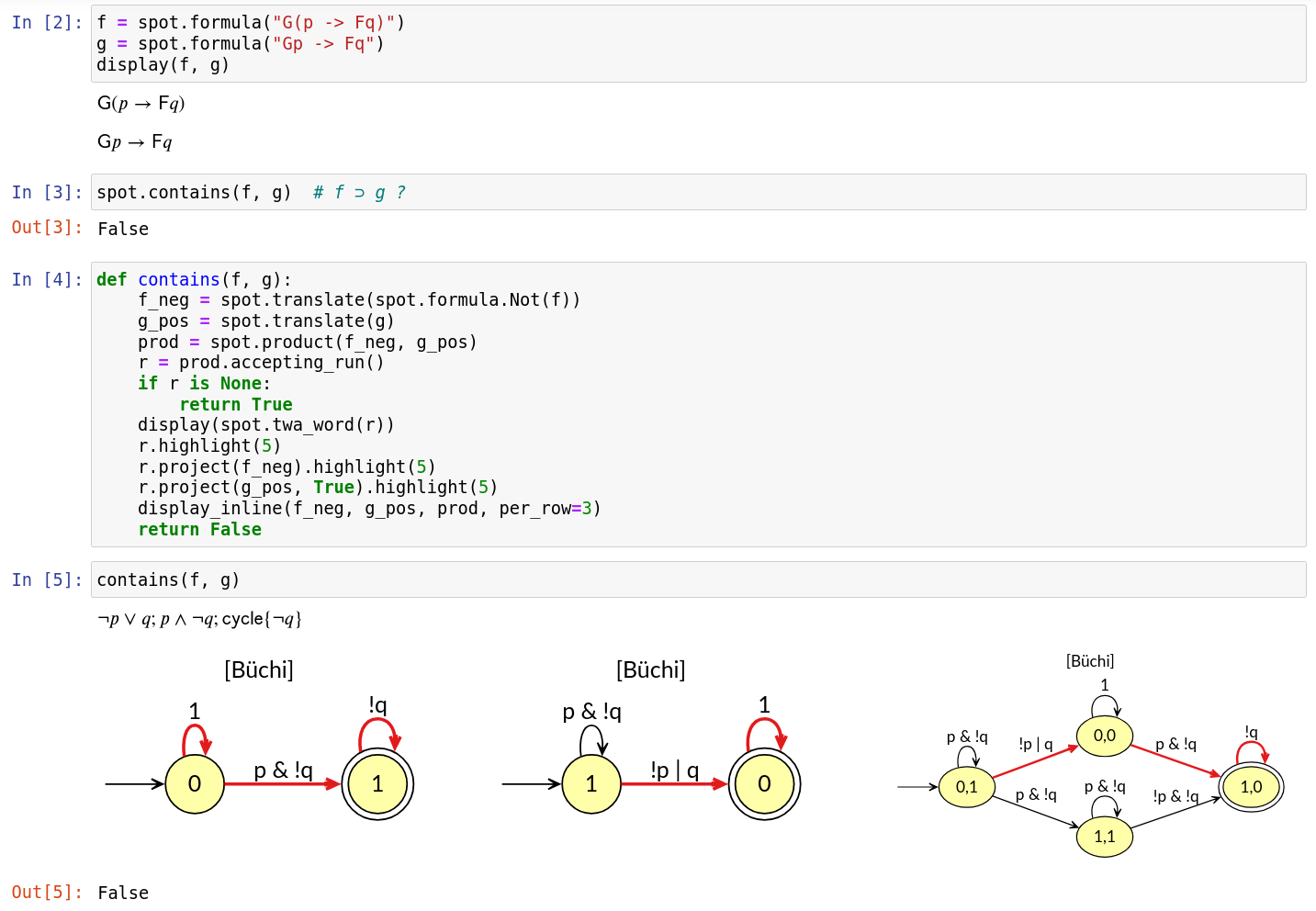}
  \caption{Using Spot in a Jupyter Notebook}
  \label{fig:notebook}
\end{figure}

Figure~\ref{fig:notebook} shows an example where the
\texttt{spot.contains()} function is first used to check inclusion
between the languages represented by two LTL formulas.  Then a
\texttt{contains()} function is defined to demonstrate how such an
inclusion check could be implemented using non-deterministic Büchi
automata, highlighting a counterexample annotated on those automata.

\begin{figure}[t]
  \centering
  \includegraphics[width=\textwidth]{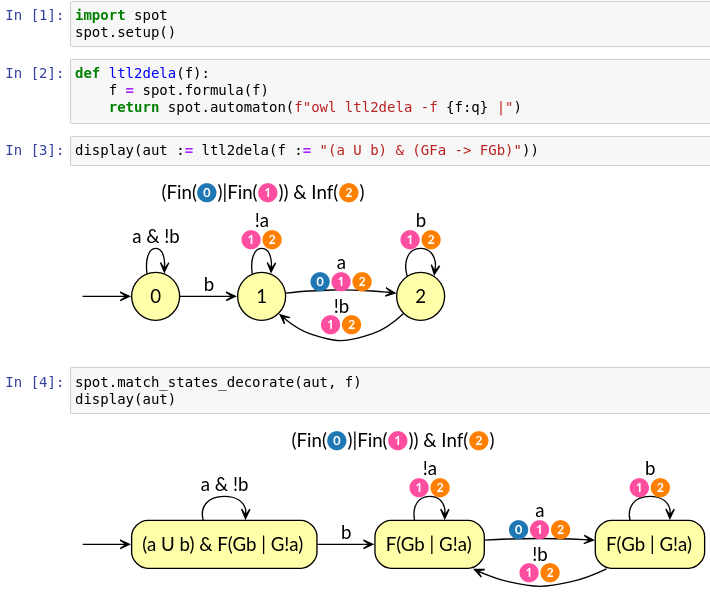}
  \caption{Running Owl's \texttt{ltl2dela} tool, and then relabeling
    the states of the produced automaton by the LTL formulas they
    recognize.}
  \label{fig:match}
\end{figure}

To help interpret automata and their connection to logic, a feature
recently introduced in Spot allows relabeling the states of a
deterministic automaton that recognizes a known formula.  Figure~\ref{fig:match}
demonstrates this on an automaton produced by Owl~\cite{kretinsky.18.owl}.

More advanced scenarios, such as LTL model
checking\footnote{\url{https://spot.lre.epita.fr/ipynb/atva16-fig2b.html}}
or LTL
synthesis\footnote{\url{https://spot.lre.epita.fr/ipynb/synthesis.html}},
can also be covered.

For courses where a local Python/Jupyter installation is impractical,
\emph{spot-sandbox}\footnote{\url{http://spot-sandbox.lre.epita.fr/}}
provides a pre-configured Jupyter instance with Spot already
installed.  Students connect through a browser and start experimenting
immediately.  The Docker image serving \emph{spot-sandbox} can also be
downloaded to be run locally.

\section{Command-line Tools}

For teaching activities, it is often useful to look for examples
satisfying certain criteria.  For instance, we might be looking for
examples of formulas that are equivalent to $a \U b$,
or some formula that belongs to a syntactic class above its
property class, or formulas that translate to automata with fewer than
3 states, etc.

Spot's command-line tools make it easy to brute-force such examples by
generating random formulas, and retaining only those matching the
criteria~\cite{duret.13.atva}. For instance, here is a list of
formulas equivalent to $a\U b$, not using $\X$ or $\mathsf{xor}$; this
command simply generates an infinite list (\texttt{-n-1}) of random
formulas, and then filters the formulas that are equivalent to
$a \U b$ and stops after 10 have been found.

\begin{lstlisting}[style=ShStyle]
% randltl --ltl-prio=X=0,xor=0 -n-1 a b | ltlfilt --equivalent-to='a U b' -n10
\end{lstlisting}
\begin{lstlisting}[style=ShStyleOut]
a U b
b W (a U b)
((a | Fb) -> b) M (a U b)
a U (Gb W b)
b U (a U b)
(b W (a & Fb)) W b
a U (b | (b & ((0 R b) R Ga)))
(a <-> Fb) U b
b M (b M (a U b))
b M (a | b)
\end{lstlisting}

The next command looks for 10 formulas that are stutter-invariant
despite containing an $\X$ operator.  The command also excludes
formulas that are equivalent to $\top$ or $\bot$ (option \texttt{-v}
negates the matches), and simplifies formulas to also ignore formulas
where the $\X$ are easy to remove.
\begin{lstlisting}[style=ShStyle]
% randltl -n-1 a b | grep X | ltlfilt --stutter-invariant | \
ltlfilt -v --equivalent-to=0 | ltlfilt -v --equivalent-to=1 | \
ltlfilt --simplify | grep X | head
\end{lstlisting}
\begin{lstlisting}[style=ShStyleOut]
b | (a & XF(b R a)) | (!a & XG(!b U !a))
G(b U (!b & X!b))
(b | X!b) W a
Xa W (b | ((!b M F!a) R !a))
!b & G(G!a | XFa)
G(!b & X(!b W XG!a))
b | (a & XF!a)
Ga | G(!a & X(!a U !b))
a | (((!b & Xb) | (b & X!b)) M !b)
GF(Ga | (b R X!b))
\end{lstlisting}

The following command looks for 10 obligation formulas that cannot be
detected as obligations by looking at their syntactic classes.  The
option \verb|--stats=%h,%f| asks \texttt{ltlfilt} to output a CSV file
where the first column is a letter indicating the class in the Manna
and Pnueli hierarchy (\texttt{O} is for \emph{obligation}), and the
second column is a formula.  After grepping for lines that start with
\texttt{O}, \texttt{ltlfilt} is used again to extract the first 10
lines of the second column (\texttt{-/2} designates the second column
of standard input).
\begin{lstlisting}[style=ShStyle]
% randltl -n-1 a b | \
ltlfilt --simplify -v --syntactic-obligation --stats=%h,%f | \
grep '^O,' | ltlfilt -F-/2 --stats=%f -n10
\end{lstlisting}
\begin{lstlisting}[style=ShStyleOut]
(a & G(XG!a | (b U Xa))) | (!a & F(XFa & (!b R X!a)))
G(b U (!b & X!b))
(b U Ga) | (Fa & Xb)
(a U b) & ((b U a) R XF!a)
XX(!b R (!a | X(!a M F!b)))
Fa R X(b & X(b | X(!b U !a)))
X(G!a | (b & X(b U G!b)))
(b U (!b U Gb)) W Xb
X((F!a W b) M XGa)
XF((G!a & XFb) | (a & X(G!b | Fa)))
\end{lstlisting}

\section{Conclusion}
\label{sec:conclusion}

Spot offers several interfaces for teaching LTL and $\omega$-automata:
a web application for quick browser-based exploration; Jupyter
notebooks that weave narrative, live code, and inline automaton
drawings; and command-line tools for easy chaining of commands with
pipes.  All three interfaces expose the same underlying library.

Spot is actively maintained under a GPL v3 license.  However, most new
features are currently driven by research needs rather than
pedagogical ones.  Instructors who find functionality missing or
workflows that could be streamlined for classroom use are warmly
encouraged to open feature requests on the issue tracker at
\url{https://gitlab.lre.epita.fr/spot/spot}, so that teaching use
cases can be better prioritized alongside research use cases.

\bibliography{teal2026-spot}

\end{document}